\documentclass[twoside]{dis08} 
\usepackage[latin1]{inputenc} 
\usepackage[dvips]{graphicx,epsfig,color} 
\usepackage{wrapfig,rotating} 
\usepackage{amssymb,amsmath,array} 
 
\begin{document} 
\title{T-odd Effects in Photon-Jet Production at the Tevatron} 
 
\author{C. Pisano, D. Boer and P.J. Mulders 
\vspace{.3cm}\\ 
Vrije Universiteit Amsterdam, Department of Physics and Astronomy\\ 
NL-1081 HV Amsterdam, the Netherlands}

 
\maketitle 
 
\begin{abstract} 
The angular distribution in photon-jet production in 
$p\,\bar{p} \rightarrow \gamma \,{\rm jet} \,X$ is 
studied within a generalized factorization scheme 
taking into account the transverse momentum of the partons in the initial 
hadrons. Within this scheme an anomalously large $\cos(2\phi)$ asymmetry 
observed in the Drell-Yan process could be attributed to 
the T-odd, spin and transverse momentum dependent 
parton distribution function $h_1^{\perp\,q}(x, \boldsymbol p_{\perp}^2)$. 
The same function is expected to produce a $\cos(2\phi)$ 
asymmetry in the photon-jet production cross section. 
This particular azimuthal asymmetry is estimated to be smaller than the 
Drell-Yan asymmetry but still of considerable size for Tevatron kinematics,  
offering a new possibility to study T-odd effects at the Tevatron. 
\end{abstract}

 
\begin{figure}[t] 
\begin{center}
\epsfig{figure=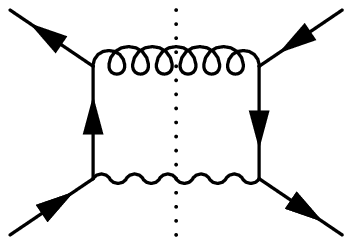,width=2.2cm}  \hspace{0.5cm} 
\epsfig{figure=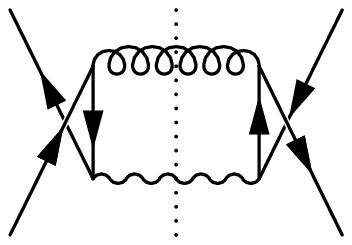,width=2.2cm}  \hspace{0.5cm} 
\epsfig{figure=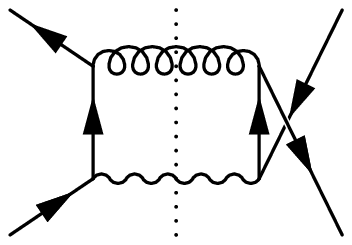,width=2.2cm}  \hspace{0.5cm} 
\epsfig{figure=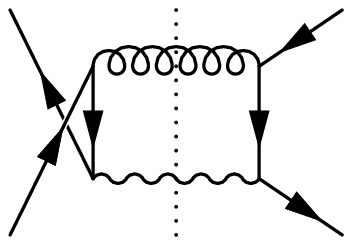,width=2.2cm} 
\caption{Cut diagrams for the subprocess $q \bar q \rightarrow \gamma g$.} 
\label{fig:qq} 
\vspace{0.8cm} 
\epsfig{figure=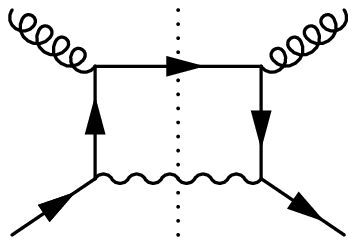,width=2.2cm}  \hspace{0.5cm} 
\epsfig{figure=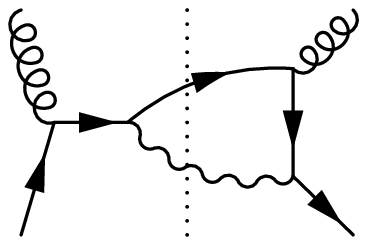,width=2.2cm}  \hspace{0.5cm} 
\epsfig{figure=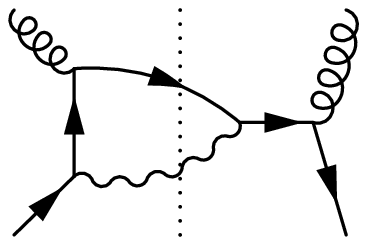,width=2.2cm}  \hspace{0.5cm} 
\epsfig{figure=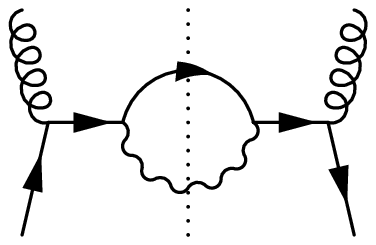,width=2.2cm} 
\caption{Cut diagrams for the subprocess $q g\rightarrow \gamma q$.} 
\end{center}
\label{fig:qg} 
\end{figure} 

In this contribution to DIS 2008~\cite{url} we consider the process:
\begin{equation} 
h_1(P_1){+}h_2(P_2)\, {\rightarrow}\,\gamma(K_\gamma){+}{\rm jet}(K_j){+}X \, , 
\end{equation} 
where the four-momenta of the particles are given within brackets, and 
the photon-jet pair in the final state is almost back-to-back in the plane perpendicular to the direction of the incoming hadrons.  
To lowest order in pQCD the reaction is described in terms of the partonic  
two-to-two subprocesses $q (p_1) + \bar{q}(p_2) \rightarrow \gamma (K_\gamma) + g(K_j)$ and  
$q (p_1) + {g}(p_2) \rightarrow \gamma (K_\gamma) + q(K_j)$. 
Following reference~\cite{Boer:2007nd}, we will instead of collinear 
factorization consider a generalized factorization scheme taking into account
partonic transverse momenta. 
We make a lightcone decomposition of the hadronic momenta in terms of two light-like Sudakov vectors $n_+$ and $n_-$, satisfying $n_+^2\,{=}\,n_-^2\,{=}\,0$ and $n_+{\cdot}n_-\,{=}\,1$: 
\begin{equation} 
P_1^\mu 
=P_1^+n_+^\mu+\frac{M_1^2}{2P_1^+}n_-^\mu\ ,\qquad\text{and}\qquad 
P_2^\mu 
=\frac{M_2^2}{2P_2^-}n_+^\mu+P_2^-n_-^\mu\ ~. 
\end{equation} 
In general $n_+$ and $n_-$ will define the lightcone components of every 
vector $a$ as $a^\pm \equiv a \cdot n_\mp$, while  
perpendicular vectors $a_\perp$  will always refer to the components of   
 $a$ orthogonal to both incoming hadronic momenta, $P_1$ and $P_2$. Hence 
the partonic momenta ($p_1$, $p_2$) can be expressed  in terms of the   
lightcone momentum fractions ($x_1$, $x_2$) and the  
intrinsic transverse momenta ($ p_{1 \perp}$, $ p_{2 \perp}$), as follows 
\begin{equation} 
p_1^\mu 
=x_1^{\phantom{+}}\!P_1^+n_+^\mu 
+\frac{m_1^2{+}\boldsymbol p_{1\perp}^2}{2x_1^{\phantom{+}}\!P_1^+}n_-^\mu 
+p_{1\perp}^\mu\ , 
\qquad\text{and}\qquad 
p_2^\mu 
=\frac{m_2^2{+}\boldsymbol p_{2\perp}^2}{2x_2^{\phantom{-}}\!P_2^-}n_+^\mu 
+x_2^{\phantom{-}}\!P_2^-n_-^\mu+p_{2\perp}^\mu\ .\label{PartonDecompositions} 
\end{equation} 
We denote with $s$ the total energy squared in the hadronic  
center-of-mass (c.m.) frame, $s =(P_1+P_2)^2 = E^2_{\rm c.m.} $, and  
 with $\eta_i$  the pseudo-rapidities  
of the outgoing particles, 
\emph{i.e.}\ $\eta_i\,{=}\,{-}\ln\big(\tan(\frac{1}{2}\theta_i)\big)$,  $\theta_i$ being the polar angles of the outgoing particles in the same frame.  
Finally, we introduce the  partonic Mandelstam variables   
$\hat s = (p_1 + p_2)^2$, $\hat t = (p_1-K_\gamma)^2 $ and  
$\hat u = (p_1-K_j)^2$,  
which satisfy the relations 
\begin{equation} 
\label{Yexpression} 
 -\frac{\hat t}{\hat s}  
\equiv y = \frac{1}{e^{\eta_\gamma -\eta_j}\,{+}\,1}~ , \qquad {\rm and} \qquad   -\frac{\hat u}{\hat s} = 1-y~. 
\end{equation} 
 
Along the lines of~\cite{Boer:2007nd,Bacchetta:2007sz},  
we assume that at sufficiently high energies the  
hadronic cross section  
factorizes in a soft parton correlator for each observed hadron and a hard  
part: 
\begin{eqnarray} 
d\sigma^{h_1 h_2 \rightarrow \gamma {\rm jet} X}   
& = &\frac{1}{2 s}\, \frac{d^3 K_\gamma}{(2\pi)^3\,2E_\gamma} 
\frac{d^3K_j}{(2\pi)^3\,2E_j} 
{\int}d x_1\, d^2\boldsymbol p_{1\perp}\,d x_2\, d^2\boldsymbol p_{2\perp}\, (2\pi)^4 
\delta^4(p_{1}{+}p_{2}{-}K_{\gamma}{-}K_{j}) 
 \nonumber \\ 
&&\qquad \times \sum_{a{,} b{,} c}\  
\Phi_a(x_1{,}p_{1\perp})\otimes\Phi_b(x_2{,}p_{2\perp}) 
\otimes\,|H_{ab\rightarrow \gamma c}(p_1, p_2, K_{\gamma}, K_j)|^2\ , 
\label{CrossSec} 
\end{eqnarray} 
where the sum runs over all the incoming and outgoing partons 
taking part in the subprocesses $q\bar q \to \gamma g$ and $q g \to \gamma q$. 
The convolutions $\otimes$ indicate the appropriate traces over Dirac  
indices and $|H|^2$ is the hard partonic squared amplitude, 
obtained from the cut diagrams in Figs.~\ref{fig:qq} and  2.  
The parton correlators $\Phi_a$ describe  
the hadron $\rightarrow$ parton transitions and their  parameterization in  
terms of transverse momentum dependent (TMD) distribution functions can be  
found, for instance, in~\cite{Boer:2007nd}.

\begin{figure}[t] 
\begin{center} 
\epsfig{figure=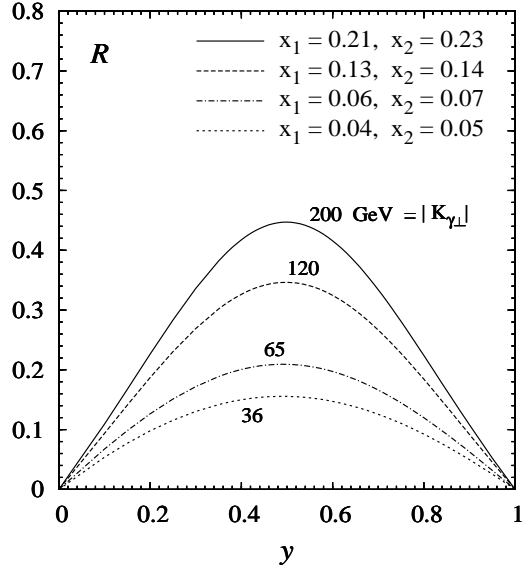, width = 8cm, height = 7.7cm} 
\caption{The ratio $R$ as a function of $y$,  
calculated according to (\ref{eq:ratio1}) and  (\ref{eq:ratio2}) for 
different values of 
$x_1$, $x_2$, $|\boldsymbol{K}_{\gamma\perp}|$ typical of the Tevatron  
experiments~\cite{Abazov:2008er}.}  
\label{fig:ratio} 
\end{center} 
\end{figure} 

We are interested in events in which the photon and the jet 
are approximately back-to-back in the transverse plane, therefore we define  
the vector $\boldsymbol{q}_\perp \equiv \boldsymbol{K}_{\gamma \perp}  
+ \boldsymbol{K}_{j \perp}$ and require that   
$|\boldsymbol{q}_\perp| \ll |\boldsymbol{K}_{\gamma\perp}|,  
|\boldsymbol{K}_{j\perp}|$. Neglecting power-suppressed terms of the order ${\cal{O}}(1/(\boldsymbol K_{\gamma\perp}^4 s))$ and using  
the approximation $|\boldsymbol{K}_{\gamma \perp}| \approx |\boldsymbol{K}_{j \perp}|$, we obtain 
\begin{eqnarray} 
\frac{d\sigma^{h_1 h_2\to \gamma {\rm jet} X}} 
{d\eta_\gamma\, d\eta_j \,d^2 \boldsymbol K_{\gamma\perp}\, d^2 \boldsymbol q_{\perp} }    
&  =  & \frac{1}{\pi^2}\,\frac{d\sigma^{h_1 h_2\to \gamma {\rm jet} X}} 
{d\eta_\gamma\, d\eta_j \,d \boldsymbol K_{\gamma\perp}^2\, d \boldsymbol q_{\perp}^2 } \,   \bigg (1 + {\cal{A}}(y, x_1, x_2, \boldsymbol{q}_\perp^2) \,  
\cos 2 (\phi_\perp -\phi_\gamma)    \bigg )\, , \nonumber \\ 
\label{eq:csfinal}  
\end{eqnarray} 
with $\phi_\gamma$ and $\phi_\perp$ being the azimuthal angles, 
in the hadronic center-of-mass frame, 
of the outgoing photon and of the vector $\boldsymbol{q}_\perp$ respectively. 
The  azimuthal asymmetry is given by  
\begin{eqnarray} 
{\cal{A}}(y, x_1, x_2, \boldsymbol{q}_\perp^2) & = &  
 \nu (x_1, x_2, \boldsymbol{q}_\perp^2) \,{R}(y, x_1, x_2, \boldsymbol{q}_\perp^2) \, ,  
\label{eq:asymmetry} 
\end{eqnarray} 
where $\nu$ contains the dependence on the time-reversal (T) odd distribution function  $h_1^{\perp\,q}(x, \boldsymbol{p}_{\perp}^2)$ and is identical to 
the azimuthal asymmetry expression that appears in the Drell-Yan process~\cite{Boer:1999mm},  
with the scale $Q$ equal to $|\boldsymbol{K}_{\gamma \perp}|$.  The function $h_1^{\perp q}(x, \boldsymbol{p}_\perp^2)$ is interpreted as the quark 
transverse spin distribution in an unpolarized hadron~\cite{Boer:1997nt}. The role of gauge links in the calculation of $\nu$ is 
discussed in detail in~\cite{Boer:2007nd}. 
The ratio $R$  only depends on the T-even unpolarized distribution functions $f_1^{q, g} (x, \boldsymbol{p}_{\perp}^2)$, which integrated over $\boldsymbol{p}_\perp$ give the familiar lightcone momentum distributions $f_1^{q, g}(x)$.  
The  explicit expressions of $\nu$ and $R$ are given in~\cite{Boer:2007nd}.
 
The process $p \,\bar{p} \rightarrow \gamma \,{\rm jet}\, X $ is currently 
being analyzed by the D\O\ Collaboration at the  Tevatron  collider. 
Data on the cross section, differential in 
$\eta_\gamma$, $\eta_j$ and $\boldsymbol{K}_{j\perp}^2$, have been taken at 
$\sqrt{s} = 1.96$ TeV~\cite{Abazov:2008er}. 
Such angular integrated measurements are only sensitive to the 
transverse momentum integrated parton distributions. 
A study of the angular dependent 
cross section in (\ref{eq:csfinal}) will provide valuable information on the 
TMD distribution function $h_1^{\perp\,q}(x, \boldsymbol{p}_\perp^2)$, if the
azimuthal asymmetry ${\cal{A}}$ turns out to be sufficiently 
sizeable in the available kinematic region. 
Model calculations~\cite{Boer:2002ju,Barone:2006ws} 
applied to the $p\,\bar p$ Drell-Yan process
have shown that the quantity $\nu$ in (\ref{eq:asymmetry})
 is of the order of 30\% or higher for $|\boldsymbol{q}_\perp|$ of a few GeV
 and $Q$ values of ${\cal O}(1-10)$ GeV. 
Therefore, a study of the order of magnitude of ${\cal A}$ 
as a function of $x_1$, $x_2$ and $\boldsymbol{q}_\perp^2$ requires 
an estimate of the ratio $R$  in (\ref{eq:asymmetry}). This will be obtained 
as follows.

First of all, the  
unknown TMD distribution functions appearing in  the definition of $R$ are 
 evaluated assuming a
factorization of their transverse momentum dependence,  that is  
$f_1^{q, g}(x_, \boldsymbol{p}_{\perp}^2) = f_1^{q, g}(x) 
{\cal T}(\boldsymbol{p}_\perp^2)$, 
with  $f_1^{q, g}(x)$ being the usual unpolarized parton distributions and 
${\cal T}(\boldsymbol{p}_\perp^2)$ being a generic function, taken to be the 
same for all partons and often chosen to be Gaussian. The
$\boldsymbol{q}_\perp^2$-dependence of $R$ then drops out and we get 
\begin{equation} 
R = \frac{2 N^2 y (1-y) \sum_q e^2_q \,f_1^q(x_1)f_1^q(x_2)}{D(y, x_1, x_2)}~,
\label{eq:ratio1}  
\end{equation} 
where $N$ is the number of colors and 
\begin{eqnarray} 
D(y, x_1, x_2) & = &  \sum_q e^2_q \big \{{N} (1-y)(1+y^2)f_1^q(x_1)f_1^g(x_2) + {N} y (1+(1-y)^2)f_1^q(x_2)f_1^g(x_1)\nonumber  \\  
&& + 2\, (N^2-1)(y^2+(1-y)^2)  f_1^q(x_1)f_1^q(x_2) \big \}. 
\label{eq:ratio2} 
\end{eqnarray} 
Here we have used that the 
antiquark contribution in the antiproton equals the quark
contribution inside a proton.  
We consider only light quarks, {\it i.e.\/} the sum in (\ref{eq:ratio2})    
runs over  
$q = u$, $\bar{u}$, $d$, $\bar{d}$, $s$, $\bar{s}$, and 
we use the leading order GRV98 set~\cite{Gluck:1998xa} for the parton 
distributions, at the scale $\mu^2 = \boldsymbol{K}_{\gamma\perp}^2$.  
 
Our results for $R$ as a function of  $y$ are shown in Fig.~\ref{fig:ratio} 
at some fixed values of the variables $x_1$, $x_2$ and $|\boldsymbol{K}_{\gamma\perp}|$, 
typical of the Tevatron experiments~\cite{Abazov:2008er}. 
The values of $x_1$ and $x_2$ considered correspond to their average when both the photon and the jet are  in the  
central rapidity region, where $\eta_j\approx \eta_{\gamma}\approx 0$ and  
$x_1\approx x_2$. 
In this case  $y\approx 0.5$, where $R$ turns out to be largest.  
Evidently,  
$R$ increases as $x_1$ and $x_2$ increase, due to the  
small contribution, in the denominator, of the gluon distributions $f_1^g(x)$  
in the valence region.   
 
Hence, we see that the asymmetry ${\cal A}$ is a product of a large  
Drell-Yan asymmetry term $\nu$ and a factor $R$ that is estimated to be in the 
10\%-50\% range for Tevatron kinematics. This leads us to conclude that an 
asymmetry ${\cal A}$ in the order of 5\%-15\%  is possible in the central  
region.   
This could allow a study of the distribution function $h_1^{\perp\,q}$ in  
$p\,\bar p \rightarrow \gamma \,{\rm jet}\, X$ at the Tevatron, offering a 
new possibility of measuring T-odd effects using a high energy collider.



\section*{Acknowledgments}
We would like to thank C. Bomhof for useful comments and discussions. 
This research is part of the 
   research program of the ``Stichting voor Fundamenteel Onderzoek der 
   Materie (FOM)'', which is financially supported by the ``Nederlandse 
   Organisatie voor Wetenschappelijk Onderzoek (NWO)''.

 
\begin{footnotesize}

\end{footnotesize}

 
\end{document}